# A Simulation Based Performance Comparison Study of Stability-Based Routing, Power-Aware Routing and Load-Balancing On-Demand Routing Protocols for Mobile Ad hoc Networks

Natarajan Meghanathan
Assistant Professor of Computer Science
Jackson State University
Jackson, MS 39217, USA
nmeghanathan@jsums.edu

Leslie C. Milton
Computer Scientist
US Army Engineer Research & Development Center
Vicksburg, MS 39180, USA
Leslie.C.Milton@usace.army.mil

*Abstract*— **The high-level contribution of this paper is a simulation-based detailed performance comparison of three different classes of on-demand routing protocols for mobile ad hoc networks: stability-based routing, power-aware routing and load-balanced routing. We choose the Flow-Oriented Routing protocol (FORP), Min-Max Battery Cost Routing (MMBCR) and the traffic interference based Load Balancing Routing (LBR) protocol as representatives of the stability-based routing, power-aware routing and load-balancing routing protocols respectively. FORP incurs the least number of route transitions; while LBR incurs the smallest hop count and lowest end-to-end delay per data packet. Energy consumed per data packet is the least for LBR, closely followed by MMBCR. FORP incurs the maximum energy consumed per data packet, both in the absence and presence of power control. Nevertheless, in the presence of power control, the end-to-end delay per data packet and energy consumed per data packet incurred by FORP are significantly reduced compared to the scenario without power control. MMBCR is the most fair in terms of node usage and incurs the largest time for first node failure. FORP tends to repeatedly use nodes lying on the stable path and hence is the most unfair of the three routing protocols. FORP also incurs the smallest value for the time of first node failure.**

## I. INTRODUCTION

A mobile ad hoc network (MANET) is a dynamic distributed system of wireless nodes that move independently and arbitrarily. MANET nodes operate with reduced battery charge and have limited transmission range. In MANETs, reactive on-demand routing protocols (that determine routes only when required) incur less overhead and exhibit better performance compared to the class of proactive routing protocols (that determine routes for every node pairs, irrespective of the requirement) [5][10]. We restrict ourselves to on-demand routing protocols in this paper.

Based on the principle and/or the metric for route selection, on-demand routing protocols can be categorized into different classes: power-aware routing, load-balanced routing, minimum-hop/delay based routing, stability-based routing and etc. Examples of minimum-hop/delay based routing protocols include the Dynamic Source Routing (DSR) [11] and Ad hoc On-demand Distance Vector (AODV) [17] routing protocols. Examples of stability-based routing protocols are the Flow-oriented Routing Protocol (FORP) [18], Associativity-based Routing (ABR) [19] and the Route-lifetime Assessment Based Routing (RABR) protocols [1]. FORP determines the sequence of most stable paths among the stability-based routing protocols [14]. Min-Max Battery Cost Routing (MMBCR) [20] is a power-aware routing algorithm proposed to maximize the time of first node failure as this routing algorithm selects routes with the objective of maximizing the minimum residual battery power of a node in the route. MMBCR can be implemented on the top of any on-demand routing protocol like DSR, AODV and etc. The load-balancing routing (LBR) protocol [8] routes data packets by circumventing congested paths and balances the traffic load to yield a lower end-to-end delay per data packet. LBR outperforms both DSR and AODV by yielding a higher packet delivery ratio and a lower end-to-end delay per data packet [8].

Transmission power control (TPC) is the technique of dynamically adjusting the transmission power of the sending node based on the distance to the intended receiving node of the packet [13]. We refer to the end nodes of a hop as sender and receiver and the end nodes of a path as source and destination. Without TPC, the transmission power per hop is fixed and is based on the transmission range of the sender node. With TPC, the transmission power spent to send a packet on a hop is a function of the distance between the sender and receiver, which is less than or equal to the transmission range of the sender node. TPC helps to reduce the energy consumed in sending a packet from a source to its destination across multiple hops. TPC also increases bandwidth usage, because we freeze (from transmission and reception at a hop) only the nodes whose distance to the sender/ receiver is less than or equal to the distance between the sender and the receiver.

Most of the performance comparison studies (e.g., [4][5][10][15]) on on-demand MANET routing protocols have been focused on the minimum-hop/delay based routing protocols. To the best of our knowledge, we could not find any work that has compared the performance of the following three different categories of routing protocols: power-aware routing protocols to maximize the time of first node failure, load-balanced routing and the stability-based routing protocols. In this work, we choose MMBCR (implemented on the top of DSR), LBR and FORP respectively to be the representatives of the power-aware routing protocols to maximize the time of first node failure, load-balanced routing and the stability-based routing categories. We implement all these three routing protocols in ns-2 [6] and study their performance with respect to several metrics, both in the absence and in the presence of TPC.

The rest of the paper is organized as follows: In Section 2, we provide a brief overview of the FORP, LBR and the MMBCR protocols. Section 3 describes the simulation environment in detail and introduces the performance metrics measured. Section 4 illustrates the performance results obtained and interprets the nature of the results for each metric. Section 5 summarizes all the performance results and presents the conclusions.

## II. REVIEW OF MANET ROUTING PROTOCOLS

This section provides a brief overview of the FORP, LBR and MMBCR protocols. We first describe a generic flooding-based route discovery approach that we use to discover routes for the above three routing protocols studied in this paper.

### A. Flooding-Based Query Reply Cycle

Whenever a source node *s* has data to send and does not know about any route to a destination node *d*, the source initiates flooding by propagating a Route-Request (RREQ) packet among its neighbors. Each intermediate node upon receiving the RREQ packet will rebroadcast the packet if the node has not seen a RREQ packet with a sequence number greater than or equal to that in the current RREQ packet. Before forwarding the RREQ packet to the neighbors, the intermediate node inserts its own ID on the RREQ packet and updates the cost field for the upstream link on which the RREQ was received. The destination receives RREQ packets along several paths and selects the path that best satisfies the route selection principles/ metric of the particular routing protocol in use. The destination sends a Route-Reply (RREP) packet on the reverse of the selected path so that the packet now travels from the destination back to the source. The destination also includes the link-wise *s-d* path information in the RREP packet. The intermediate nodes on the selected path learn about their inclusion in the *s-d* path after receiving the RREP packet. The source starts transmitting the data packet on the *s-d* path learnt from the RREP packet. When an intermediate node on an *s-d* path cannot forward the data packet to a downstream node, the intermediate node sends a Route-Error (RERR) packet to the source, which initiates a new flooding-based route discovery.

### B. Flow-Oriented Routing Protocol (FORP)

FORP [18] utilizes the mobility and location information of the nodes to approximately predict the expiration time (LET) of a wireless link. The minimum of LET values of all wireless links on a path is termed as the route expiration time (RET). The route with the maximum RET value is selected. FORP assumes the availability of location-update mechanisms like GPS (Global Positioning System) [9] to identify the location of nodes and also assumes that the clocks across all nodes are synchronized. Each node is assumed to be able to predict the LET values of each of its links with neighboring nodes based on the location and mobility information exchanged periodically in the neighborhood. Route discovery is similar to the flooding-based query-reply cycle described in Section 2.1, with the information propagated in the RREQ packet being the predicted LET of each link in a path.

Let two nodes *i* and *j* be within the transmission range of each other. Let $(x_i, y_i)$ and $(x_j, y_j)$ be the co-ordinates of the mobile hosts *i* and *j* respectively. Let $v_i$, $v_j$ be the velocities and $\Theta_i$, $\Theta_j$, where $(0 \leq \Theta_i, \Theta_j < 2\pi)$ indicate the direction of motion of nodes *i* and *j* respectively. The amount of time the two nodes *i* and *j* will stay connected, $D_{i-j}$, can be predicted using the following equation:

$$D_{i-j} = \frac{-(ab+cd) + \sqrt{(a^2+c^2)r^2 - (ad-bc)^2}}{a^2+c^2}$$

where: $a = v_i \cos\Theta_i - v_j \cos\Theta_j$; $b = x_i - x_j$; $c = v_i \sin\Theta_i - v_j \sin\Theta_j$; $d = y_i - y_j$

### C. Load-Balancing Routing (LBR) Protocol

The LBR protocol [8] uses the concepts of "node activity" and "traffic interference" to select the best *s-d* path that would encounter the minimum traffic load for transmission and minimum interference by neighboring nodes. Each node includes in the beacon packets information about the number of *s-d* sessions the node is part of. The activity of a node is defined as the number of active *s-d* paths (*s-d* paths that currently use the node as one of the intermediate forwarding nodes) the node is part of. The traffic interference at a node is the sum of all the activities of the neighbors of the node. For a given source *s* and destination *d*, LBR chooses an *s-d* path that has the minimum value for the sum of the activities of the intermediate forwarding nodes on the path and the traffic interferences due to the neighboring nodes of the intermediate nodes. The route selection metrics recorded in the RREQ packets are the activity and traffic interference of each of the intermediate forwarding nodes of the RREQ packet.

### D. Min-Max Battery Cost Routing (MMBCR)

The residual battery charge of an *s-d* path is the minimum of the battery charges of the intermediate nodes of the path. The MMBCR algorithm [20] chooses the *s-d* path with the largest residual battery charge. Each node periodically broadcasts a beacon packet containing information about the current battery charge available at the node. The route selection metric recorded in an *s-d* path is the residual (available) battery charge of each of the intermediate nodes on the *s-d* path through which the RREQ packet got forwarded.

## III. SIMULATION CONDITIONS AND PERFORMANCE METRICS

We use ns-2 (version 2.28) [6] as the simulator for our study. We implemented the FORP, LBR protocols and the MMBCR algorithm on top of DSR in ns-2. The network dimensions are 1000m x 1000m. The transmission range of each node is 250m. We vary the network density by conducting simulations with 50 nodes (low density network;

average of 10 neighbors per node) and 100 nodes (high density network; average of 20 neighbors per node). We conduct two sets of experiments. In the first set of experiments (Fig. 1 through 10), the energy level at each node is 1500 Joules and simulations were run for 1000 seconds. In the second set of experiments, the energy level at each node is 100 Joules and simulations were run until the time of first node failure (Fig. 11 and 12).

Traffic sources are continuous bit rate (CBR). Number of source-destination (*s-d*) sessions used is 15 (low traffic load) and 30 (high traffic load). The starting times of the *s-d* sessions is uniformly distributed between 1 to 40 seconds. Data packets are 512 bytes in size; the packet sending rate is 4 data packets per second.

The MAC layer uses the distributed co-ordination function (DCF) of the IEEE Standard 802.11 [3] for wireless LANs. With TPC, the transmission power used is calculated using the formula [14][16]: $1.1182 + 7.2 * 10^{-11}(d)^4$, which includes power required to drive the circuit (1.1182W) and transmission power from the antenna computed using the two-ray ground reflection model [6] and distance *d* between the sender and receiver estimated based on the signal strengths of the Request-to-Send (RTS) and Clear-to-Send (CTS) packets. For simulations without TPC, the fixed transmission power per hop is 1.4W. The reception power per hop is fixed for all situations and it is 0.967W.

In the presence of overhearing, no real optimization in the energy consumption or node lifetime can be achieved [12]. Thus, in this paper, we do not consider the energy lost in the idle state and focus only on the energy consumed during the transmission and reception of messages (the DATA packets, the MAC layer RTS-CTS-ACK packets and the periodic beacons) and the energy consumed due to route discoveries. We model the energy consumed due to broadcast traffic and point-to-point traffic as linear functions (like in [7]) of the packet transmission time, network density, transmission and reception powers per hop.

The node mobility model used is the Random Waypoint model [2], a widely used mobility model in MANET simulation studies. Here, each node starts moving from an arbitrary location to a randomly selected destination location at a speed uniformly distributed in the range $[0,\ldots,v_{max}]$. Once the destination is reached, the node continues to move by choosing a different target location and a different velocity. The $v_{max}$ values used are 5, 10 and 20 m/s (representing low node mobility scenarios) and 30, 40 and 50m/s (representing high node mobility scenarios).

We study the following performance metrics for the three routing protocols:

*1) Number of route transitions*: the average of the number of route discoveries per *s-d* session, averaged over all the *s-d* sessions of a simulation.

*2) Hop count per route*: average of the number of hops in the routes of an *s-d* session, time-averaged considering the duration of the paths and their hop count for all *s-d* sessions.

*3) End-to-end delay per data packet*: average of the delay incurred by the data packets that originate at the source and delivered at the destination. The delay incurred by a data packet includes all the possible delays – the buffering delay due to the route acquisition latency, the queuing delay at the interface queue to access the medium, transmission delay, propagation delay, and the retransmission delays due to the MAC layer collisions.

*4) Energy consumed per data packet*: average of the energy consumed by all the packets that originate at the source and delivered at the destination. We include the energy consumed due to transmission and reception of data packets, MAC layer packets and the energy consumed due to route discoveries.

*5) Fairness of node usage*: measured using the standard deviation of the energy consumed per node, which is the square root of the average of the squares of the difference between the energy consumed at each node and the average energy consumed per node. Ideally, the value of this metric should be zero to indicate that all nodes have been used fairly and no node is overused.

*6) Time of first node failure*: The time of first node failure due to exhaustion of battery charge during the simulation with a particular routing protocol.

IV. SIMULATION RESULTS

Each data point in Fig. 1 through 12 is an average of data collected using 5 mobility trace files and 5 sets of randomly selected 15 and 30 *s-d* sessions.

*A. Number of Route Transitions*

For all the simulation conditions (refer Fig. 1 and 2), FORP incurs the least number of route transitions; the number of route transitions incurred by MMBCR is 15 to 25% more than that of LBR. For a given network density and node mobility, there is no appreciable change in the number of route transitions, as we increase the offered data traffic load from low to high. This is because the *s-d* sessions are independent of each other. There is no significant change in the number of route transitions incurred by the three routing protocols when operated with and without TPC.

The number of route transitions incurred by FORP is the minimum because it is a stable path routing protocol and chooses the route that has the largest predicted lifetime since the time of selection. The number of route transitions incurred by FORP in high-density networks is often greater than that incurred in low-density networks by a factor of 5 to 25%. In high-density networks, as the number of nodes within the neighborhood is increased, FORP gets more chances of finding stable links with longer predicted lifetime.

*B. Hop Count per Path*

For all the simulation conditions (refer Figures 3 and 4), LBR incurs the minimum number of hops, closely followed by MMBCR. FORP incurs the maximum number of hops for all the simulation conditions. The hop count per path is not affected by TPC. Also, for a fixed network density, the hop count per path is not much affected by the offered data traffic load as the *s-d* sessions are independent of each other.

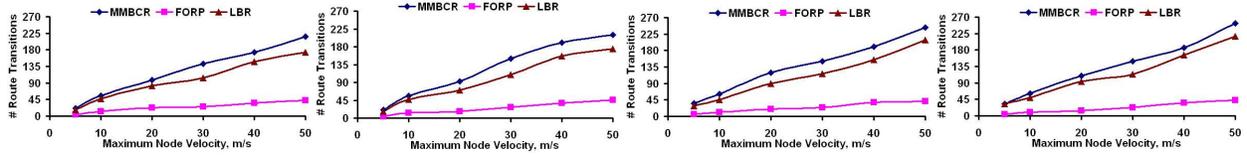

Figure 1.1. 50 nodes, 15 *s-d* pairs    Figure 1.2. 50 nodes, 30 *s-d* pairs    Figure 1.3. 100 nodes, 15 *s-d* pairs    Figure 1.4. 100 nodes, 30 *s-d* pairs

Figure 1. Route Transitions in the Absence of Transmission Power Control

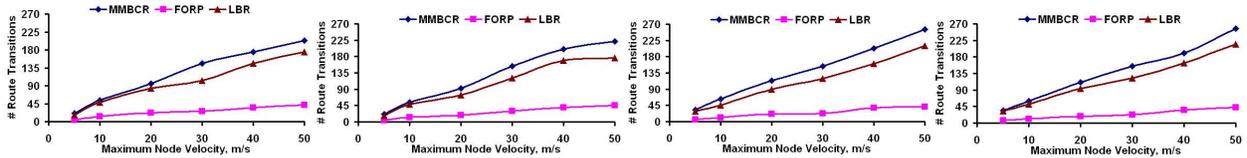

Figure 2.1. 50 nodes, 15 *s-d* pairs    Figure 2.2. 50 nodes, 30 *s-d* pairs    Figure 2.3. 100 nodes, 15 *s-d* pairs    Figure 2.4. 100 nodes, 30 *s-d* pairs

Figure 2. Route Transitions in the Presence of Transmission Power Control

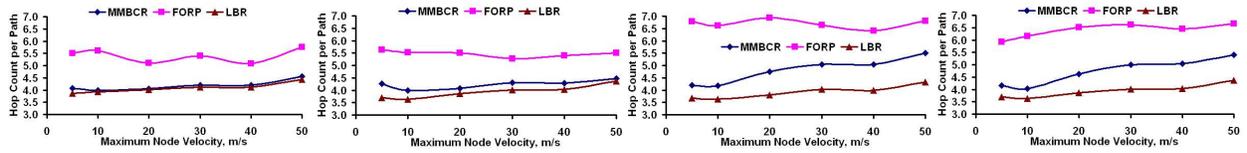

Figure 3.1. 50 nodes, 15 *s-d* pairs    Figure 3.2. 50 nodes, 30 *s-d* pairs    Figure 3.3. 100 nodes, 15 *s-d* pairs    Figure 3.4. 100 nodes, 30 *s-d* pairs

Figure 3. Average Hop Count per Path in the Absence of Transmission Power Control

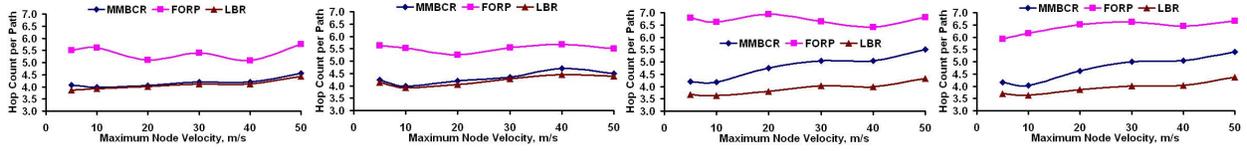

Figure 4.1. 50 nodes, 15 *s-d* pairs    Figure 4.2. 50 nodes, 30 *s-d* pairs    Figure 4.3. 100 nodes, 15 *s-d* pairs    Figure 4.4. 100 nodes, 30 *s-d* pairs

Figure 4. Average Hop Count per Path in the Presence of Transmission Power Control

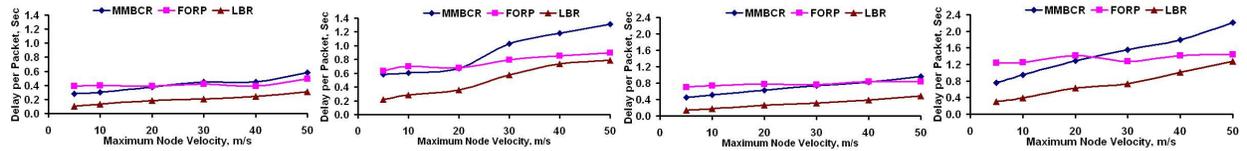

Figure 5.1. 50 nodes, 15 *s-d* pairs    Figure 5.2. 50 nodes, 30 *s-d* pairs    Figure 5.3. 100 nodes, 15 *s-d* pairs    Figure 5.4. 100 nodes, 30 *s-d* pairs

Figure 5. End-to-End Delay per Data Packet in the Absence of Transmission Power Control

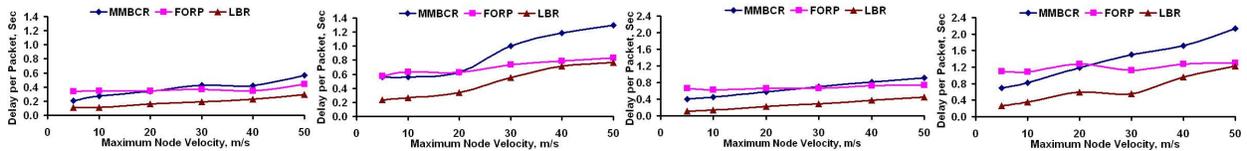

Figure 6.1. 50 nodes, 15 *s-d* pairs    Figure 6.2. 50 nodes, 30 *s-d* pairs    Figure 6.3. 100 nodes, 15 *s-d* pairs    Figure 6.4. 100 nodes, 30 *s-d* pairs

Figure 6. End-to-End Delay per Data Packet in the Presence of Transmission Power Control

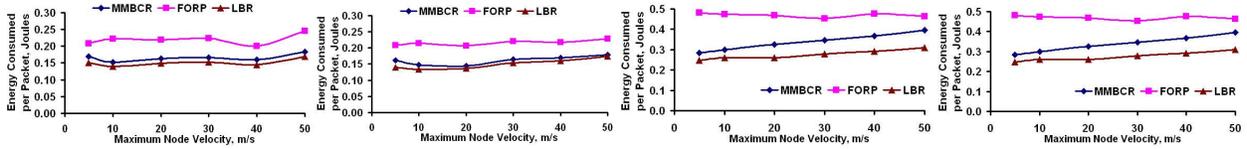

Figure 7.1. 50 nodes, 15 *s-d* pairs    Figure 7.2. 50 nodes, 30 *s-d* pairs    Figure 7.3. 100 nodes, 15 *s-d* pairs    Figure 7.4. 100 nodes, 30 *s-d* pairs

Figure 7. Energy Consumed per Data Packet in the Absence of Transmission Power Control

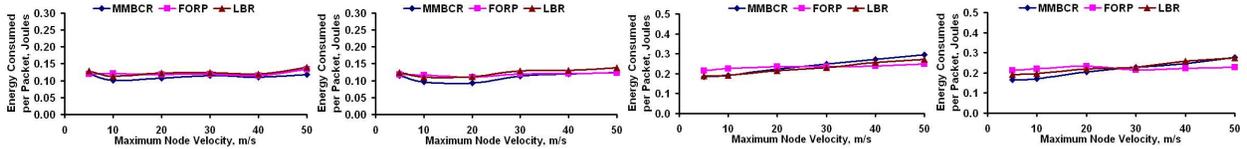

Figure 8.1. 50 nodes, 15 *s-d* pairs    Figure 8.2. 50 nodes, 30 *s-d* pairs    Figure 8.3. 100 nodes, 15 *s-d* pairs    Figure 8.4. 100 nodes, 30 *s-d* pairs

Figure 8. Energy Consumed per Data Packet in the Presence of Transmission Power Control

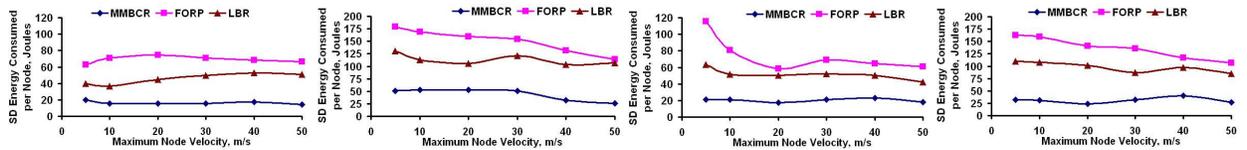

Figure 9.1. 50 nodes, 15 *s-d* pairs    Figure 9.2. 50 nodes, 30 *s-d* pairs    Figure 9.3. 100 nodes, 15 *s-d* pairs    Figure 9.4. 100 nodes, 30 *s-d* pairs

Figure 9. Standard Deviation of the Energy Consumed per Node in the Absence of Transmission Power Control

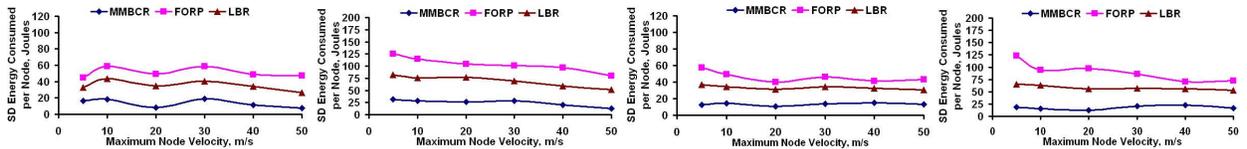

Figure 10.1. 50 nodes, 15 *s-d* pairs    Figure 10.2. 50 nodes, 30 *s-d* pairs    Figure 10.3. 100 nodes, 15 *s-d* pairs    Figure 10.4. 100 nodes, 30 *s-d* pairs

Figure 10. Standard Deviation of the Energy Consumed per Node in the Presence of Transmission Power Control

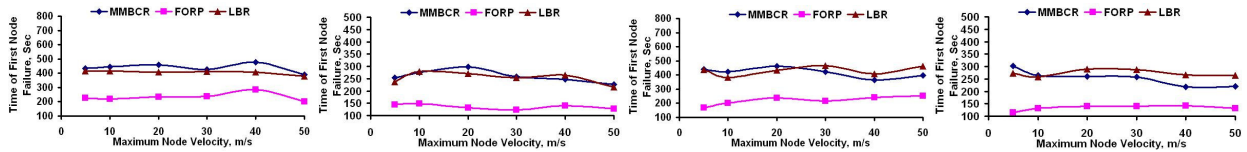

Figure 11.1. 50 nodes, 15 *s-d* pairs    Figure 11.2. 50 nodes, 30 *s-d* pairs    Figure 11.3. 100 nodes, 15 *s-d* pairs    Figure 11.4. 100 nodes, 30 *s-d* pairs

Figure 11. Time of First Node Failure in the Absence of Transmission Power Control

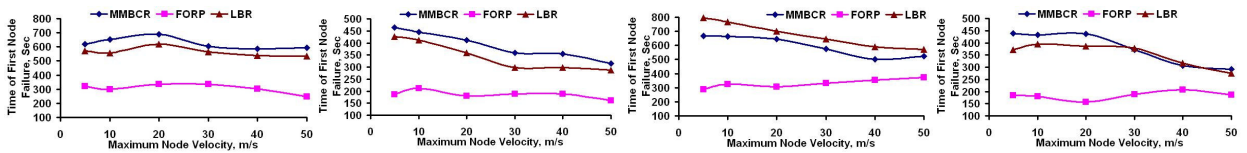

Figure 12.1. 50 nodes, 15 *s-d* pairs    Figure 12.2. 50 nodes, 30 *s-d* pairs    Figure 12.3. 100 nodes, 15 *s-d* pairs    Figure 12.4. 100 nodes, 30 *s-d* pairs

Figure 12. Time of First Node Failure in the Presence of Transmission Power Control

LBR minimizes the sum of traffic interferences of the constituent intermediate nodes of a route and this helps to minimize the number of intermediate nodes that form the route. For the source and destination of a route, the traffic interference value is considered to be zero as these nodes have to transmit and receive the data packets at any cost. MMBCR maximizes the bottleneck battery charge of a route. It attempts to avoid intermediate nodes having a lower bottleneck battery charge. The battery charge available at the source and destination nodes is not considered while choosing the bottleneck path battery charge. This indirectly helps MMBCR to find a path that has lower hop count.

As the distance separating the constituent nodes of a link is reduced, the probability of the link to have a larger predicted lifetime increases. As FORP prefers to connect the source and destination nodes using links with higher predicted lifetime (such links have shorter physical distance), the hop count of FORP paths is higher. The hop count of FORP paths in high-density networks is greater than that incurred for low-density networks by 15 to 25%. As we increase the network density, the number of nodes in a neighborhood increases and FORP gets a larger pool of links to choose from. In networks of higher density, FORP manages to find more stable links, but the hop count increases accordingly. The hop count of FORP paths is 30 to 40% and 60 to 70% of the hop count of LBR paths in networks of low density and high density respectively. The above results indicate a clear tradeoff between route stability and hop count. FORP incurs the minimum number of route transitions and MMBCR incurs the maximum number of route transitions, closely followed by LBR. On the other hand, LBR and MMBCR incur lower hop count compared to that of FORP.

### C. End-to-end Delay per Data Packet

Figures 5 and 6 indicate that the end-to-end delay per data packet for LBR is the lowest among the three protocols simulated. LBR attempts to minimize the sum of the traffic interferences of the nodes in a path; routes packets through nodes that are least congested and through the minimum number of intermediate nodes. The end-to-end delay per data packet for MMBCR is smaller than that of FORP under low node mobility conditions and found to be usually larger than that of FORP under high node mobility conditions. This is due to the significant increase in the number of MMBCR route transitions and also an appreciable increase in the hop count per path under high node mobility conditions.

The end-to-end delay per data packet incurred by FORP is the most influenced due to TPC. FORP paths have a larger hop count, but the physical length of each hop is smaller. TPC helps to reduce the forwarding and receiving interference of the data traffic load at nodes that are outside the radius of the hop length but within the transmission range of the nodes. Since there are more hops in FORP routes, the difference in the end-to-end delay per data packet in the presence and absence of TPC is very much noticeable. For a given simulation condition, the end-to-end delay per data packet for FORP in the absence of TPC is about 10-15% than that incurred in the presence of TPC. With LBR, the influence of TPC is observed in networks of high density. The reduction in the end-to-end delay per data packet for LBR and MMBCR in the presence of TPC is by 10-15% at low node mobility and below 10% at high node mobility. The reduction in the queuing delay per data packet at high node mobility for LBR and MMBCR is offset by the increase in the route-acquisition delay and the frequent route discoveries.

### D. Energy Consumed per Data Packet

Without TPC (refer Figure 7), the energy consumed per data packet is the least for LBR, followed by MMBCR. FORP incurs the maximum energy consumption per data packet as the energy consumed per data packet is directly proportional to the number of hops traversed by the data packet and for a given network density, the energy consumed per data packet is independent of the number of *s-d* sessions. For a given offered data traffic load, the energy consumed per data packet for each of the three routing protocols approximately doubles as the network density is doubled.

With TPC (refer Figure 8), the energy consumed per data packet for FORP, MMBCR and LBR is respectively about 50%, 70% and 83% of that incurred without TPC. The energy consumed per data packet is the least reduced for LBR because of the physical length of its hops being close to the transmission range of the nodes. On the other hand, for FORP, the physical length of the hops is about 50 to 60% of the transmission range of the nodes. Hence, we find relatively higher effectiveness while using TPC for FORP.

For a given offered data traffic load, the energy consumed per data packet for each of MMBCR and LBR at maximum node velocity of 5m/s is about 10% and 30 to 40% (without TPC) and 10% and 45% to 60% (with TPC) more than that incurred by these two routing protocols at maximum node velocity of 50 m/s in networks of low and high density respectively. The energy consumed per data packet for FORP is almost the same for both low and high node mobility. FORP is hence the most scalable routing protocol, both in the absence and presence of TPC, with respect to the energy consumed per data packet and node mobility.

### E. Fairness of Node Usage

Figures 9 and 10 illustrate that MMBCR is the most fair among the three routing protocols. FORP is the most unfair of all the three and LBR is in between. The fairness of MMBCR is justified by the fact that it attempts to divert routes from heavily used nodes (in terms of energy consumed) towards lightly used nodes. MMBCR chooses routes such that the bottleneck battery charge of the route is the maximum. So, MMBCR cleverly avoids from over-utilizing nodes, when there are nodes that are under-utilized. FORP incurs the highest standard deviation of energy usage because, stable paths tend to exist for a long time and use certain set of nodes preferentially over other nodes.

For all the three routing protocols, the standard deviation of energy consumed per node decreases because of TPC. This is due to the relatively lower energy consumed at all the nodes using TPC when compared with the energy consumed in the absence of TPC. Thus, the fairness of node usage of the routing protocols improves with TPC for both low and high density networks. With respect to the fairness of node usage, MMBCR gets the maximum benefit from using TPC and

FORP gets a relatively lower benefit from using TPC. The standard deviation of energy consumed per node for all the three routing decreases with increase in node mobility. When the network is highly mobile, the data forwarding load gets well-distributed among all the nodes because of frequent route transitions.

*F. Time of First Node Failure*

The time of first node failure (refer Figures 11 and 12) is high for both LBR and MMBCR and close enough to each other for most of the simulation conditions. This is due to the design of these routing protocols to prefer nodes that have been under-utilized over nodes that have been over-utilized. The time of first node failure is low for FORP in all the cases. This is due to the preferential usage of nodes lying on the stable path and the larger hop count per path. For a given offered data traffic load and network density, the time of first node failure for FORP is about 57% and 47% of the time of first node failure for MMBCR in the absence and presence of TPC respectively. FORP is not much influenced by the use of TPC at high node mobility. The increase in the number of route transitions for FORP is the lowest with increase in node mobility and the protocol spends most of the energy in transferring the data packets through stable paths of larger hop count.

All the three routing protocols exhibit an improvement in the time of first node failure while using TPC. For a given network density and offered data traffic load, the time of first node failure for all the three routing protocols in the presence of TPC is about, on average, 1.4 times to that incurred in the absence of TPC. For all the three routing protocols, there is no significant improvement in the time of first node failure, as we double the network density for a given traffic load, both in the presence and in the absence of TPC.

## V. CONCLUSIONS

The high-level contribution of this paper is a simulation-based performance comparison analysis of three different categories of MANET routing protocols: Stability-based FORP, Power-aware MMBCR and the Load-balancing routing (LBR) protocol. The simulations have been conducted under different scenarios of node density, node mobility, offered data traffic load and in the presence/absence of power control. Some of the key conclusions on the performance results are summarized below:

We observe a tradeoff between stability and hop count. FORP incurs the least number of route transitions among the three routing protocols for all the simulation conditions. MMBCR incurs the maximum number of route transitions, closely followed by LBR. LBR incurs the minimum number of hops per path, closely followed by MMBCR. FORP incurs the maximum number of hops per path for all the simulation conditions. FORP routes are more stable in networks of higher density compared to networks of lower density. The tradeoff is the increase in hop count in networks of higher density compared to those in lower density.

For a given network density, there is no significant difference in the number of route transitions and hop count per path for each of the three routing protocols when operated with and without power control and with increase in the offered data traffic load from 15 *s-d* pairs to 30 *s-d* pairs.

LBR incurs the lowest end-to-end delay per data packet among the three routing protocols for all the simulation conditions tested. The end-to-end delay per data packet for MMBCR is smaller than that of FORP under low node mobility conditions and larger than that of FORP under high node mobility conditions. The end-to-end delay per data packet incurred by FORP is the most influenced by power control. When power control is conducted on hops with smaller physical length, we manage to reduce the forwarding and receiving interference of the data traffic load at the non-participating nodes to a maximum. For a given network density and offered data traffic load, as we increase node mobility, FORP has the slowest increase in the end-to-end delay per data packet. For low and high network density, FORP and LBR are respectively the most scalable with respect to end-to-end delay per data packet as we increase the offered data traffic load. For low and high offered data traffic load, LBR and FORP are respectively the most scalable with respect to end-to-end delay per data packet as we increase the network density.

In the absence of power control, for a given simulation condition, the energy consumed per data packet is the least for LBR, followed by MMBCR. FORP incurs the maximum energy consumption per data packet in the absence of power control. On the other hand, FORP incurs the maximum reduction in the energy consumed per data packet and energy consumed per node when operated with power control. LBR incurs the least reduction in energy consumed per data packet when operated with power control. For a given offered data traffic load and network density, FORP is the most scalable routing protocol, both in the absence and presence of power control, with respect to the increase in the energy consumed per data packet and energy consumed per node, as we increase the node mobility. In the absence of power control, MMBCR incurs the least energy consumed per node, closely followed by LBR. FORP incurs the maximum energy consumption per node.

In terms of fairness of node usage, MMBCR is the most fair as it attempts to divert routes from nodes that have lost more battery charge towards nodes that have not lost relatively significant battery charge. FORP is the worst in terms of fairness of node usage as stable paths tend to exist for a long time and certain nodes are used more preferentially than others. LBR only manages to divert traffic from nodes that are currently part of multiple *s-d* sessions towards nodes that have been part of few *s-d* sessions. This strategy of LBR does not help much in shielding nodes that have lost significant battery charge. LBR prefers to route traffic through nodes which have not been forwarding much traffic, irrespective of their energy level at the nodes. This is also the reason, why LBR has slightly larger energy consumption per node than that of MMBCR, even though LBR incurs lower energy consumption per data packet than MMBCR.

The time of first node failure for both LBR and MMBCR are high and close enough to each other, while the time of first node failure for FORP is lower. The time of first node failure for each of these routing protocols in the presence of power control, is on average, 1.4 times to that incurred in the absence of power control. For all the three routing protocols, for a given offered data traffic load, there is no significant improvement in the time of first node failure, as we double the network density, both in the presence as well as in the absence of power control. MMBCR and LBR make maximum use of power control by conserving the battery charge at the nodes while transferring data packets and use that conserved energy for route discovery at high node mobility. FORP is not much influenced by the use of power control at high node mobility as it spends most of the energy in transferring the data packets through paths of larger hop count.